\documentclass[manuscript, dvipdfmx]{aastex}
\usepackage{amsmath,amssymb,subfigure}
\usepackage{color}
\usepackage{graphicx}

\newcommand{\average}[1]{\ensuremath{\langle#1\rangle} }

\begin{document}

\title{Population synthesis of planet formation using a torque formula with dynamic effects}

\author{Takanori Sasaki$^1$ and Toshikazu Ebisuzaki$^2$}
\affil{$^1$Department of Astronomy, Kyoto University, Kitashirakawa-Oiwake-cho, Sakyo-ku, Kyoto 606-8502, Japan}
\affil{$^2$RIKEN, 2-1 Hirosawa, Wako, Saitama 351-0198, Japan}
\email{takanori@kusastro.kyoto-u.ac.jp}

\begin{abstract}

Population synthesis studies into planet formation have suggested that distributions consistent with observations can only be reproduced if the actual Type I migration timescale is at least an order of magnitude longer than that deduced from linear theories. Although past studies considered the effect of the Type I migration of protoplanetary embryos, in most cases they used a conventional formula based on static torques in isothermal disks, and employed a reduction factor to account for uncertainty in the mechanism details. However, in addition to static torques, a migrating planet experiences dynamic torques that are proportional to the migration rate. These dynamic torques can impact on planet migration and predicted planetary populations. In this study, we derived a new torque formula for Type I migration by taking into account dynamic corrections. This formula was used to perform population synthesis simulations with and without the effect of dynamic torques. In many cases, inward migration was slowed significantly by the dynamic effects. For the static torque case, gas giant formation was effectively suppressed by Type I migration; however, when dynamic effects were considered, a substantial fraction of cores survived and grew into gas giants.

\end{abstract}


\newpage

\section{Introduction}

The population of extrasolar planets \citep{bar13}, and perhaps even the solar system \citep{wal11}, provide strong evidence that migration has played a role in shaping planetary systems. Low-mass planets (i.e., those with masses up to that of Neptune) migrate through the excitation of linear density waves in the disk, and through a contribution from the corotation region (i.e., Type I migration). Early analytical work \citep{tan02} focused on isothermal disks, in which the temperature was prescribed and fixed. These studies found that migration was always directed inward for reasonable disk parameters, and that migration time scales were much shorter than the disk life time; therefore, according to migration theory, all the planets should end up very close to the central star.

While Type I migration has always been linked to linear interactions with the disk, \cite{paa09} showed that corotation torque (or horseshoe drag) in isothermal disks show nonlinear behavior and can be much larger than previous linear estimates, which works against fast inward migration. However,the corotation tends to be prone to saturation and fail to prevent rapid inward migration for most of the cases, since in the absence of a diffusive process, the corotation region is a closed system; therefore, it can only provide a finite amount of angular momentum to a planet.

Several well-established theoretical models of planet formation based on the core accretion scenario adopted a population synthesis approach \citep[e.g.,][]{ida04, ida08, ida13, mor09a, mor09b}. \cite{ida04} focused on the influence of Type I migration on planetary formation processes and found that when the effects of Type I migration are taken into account, planetary cores have a tendency to migrate into their host stars before they acquire adequate mass to initiate efficient gas accretion. In order to preserve a sufficient fraction of gas giants around solar-type stars, they introduced a Type I migration reduction factor, where factor magnitudes of smaller than unity work to lengthen the Type I migration timescale relative to those deduced from linear theories. With a range of small factors ($\sim$ 0.01), it was possible to produce a planetary $M_p$-$a$ distribution that was qualitatively consistent with observations from a radial velocity survey. While several suppression mechanisms for Type I migration under various circumstances have been suggested \citep[e.g.,][]{paa11}, the origin of the extremely small reduction factor values remains unknown.

Recently, it was proposed that dynamic corotation torque can also play a role for low-mass planets, especially where static corotation torques saturate. \cite{paa14} presented an analysis of the torques on migrating, low-mass planets in locally isothermal disks. They found that planets experience dynamic torques whenever there is a radial gradient in vortensity in addition to static torques, which do not depend on the migration rate. These dynamic torques are proportional to the migration rate and can have either a positive or a negative feedback on migration, depending on whether the planet is migrating with or against the static corotation torque. Moreover, they showed that in disks a few times more massive than the minimum mass solar nebula (MMSN), the effects of dynamic torques are significant to reduce inward migration.

In this study, we deduced a torque formula for Type I migration by taking into account dynamic corrections. Using this formula, we performed population synthesis simulations with and without the dynamic corrections in order to evaluate the migration velocity quantitatively. We found that the effective torques with dynamic correction were much smaller than the simple static torques when applied to disks of the MMSN model. We used dynamic torques based on the theory of \cite{paa14} and estimated actual Type I migration, and then simulated various sets of planetary systems based on the observed range of disk properties. Finally, we compared the simulated results with observational data.

\section{Dynamic Correction of Type I Migration Formula}

According to Paardekooper et al. (2011; hereafter Pa11) and Colman \& Nelson (2014; hereafter CN14), new static torque formula for Type I migration can be derived (see Appendix). When developing the dynamic correction formula, we considered the work of \cite{paa14}, who showed that $\Gamma_{\rm dynamic}$, the term proportional to $dr_{\rm p}/dt$, must be included in the torque formula, or in other words: 
\begin{equation}
\Gamma=\Gamma_{\rm static}+\Gamma_{\rm dynamic},^{•}
\end{equation}
which is given by (Pa14 eq18):
\begin{equation}
\Gamma_{\rm dynamic}=2\pi(1-w_{\rm c}/w(r_{\rm p}))\Sigma r_{p}^2x_{\rm s}\Omega v_{\rm p},
\label{eq:TorqueDynamic}
\end{equation}
where $\Sigma$ is the surface density of the disk, $r_p$ is the semimajor axis of the protoplanet, $x_{\rm s}$ is the thickness of the horseshoe region, $\Omega=(GM_{*}/r_{\rm p}^3)^{1/2}$ ($G$ is the gravitational constant, and $M_{*}$ is the stellar mass), and $v_{\rm p}=dr_{\rm p}/dt$ is the radial velocity of the protoplanet. Here, $1-w_{\rm c}/w(r_{\rm p})$ was calculated by (Pa14 eq28, modified by TE):
\begin{equation}
1-w_{\rm c}/w(r_{\rm p})=(3/2 + p)\min\left(1, \frac{x_{\rm s}^2}{6r_{\rm p}\nu}v_{\rm p}\right),
\label{eq:Vortencity}
\end{equation}
where $p=d\ln \Sigma/d\ln r$ and $\nu$ is the viscosity of the disk. Assuming a circular orbit of the protoplanet, $v_{\rm p}$ can be calculated by the equation:
\begin{equation}
\tau_{\rm lib}\frac{dv_{\rm p}}{dt}=-v_{\rm p}+\frac{2q_{\rm d}}{\pi qr_{\rm p}^3\Omega \Sigma}(\Gamma_{\rm static}+\Gamma_{\rm dynamic}),
\label{eq:Velocity}
\end{equation}
where $\tau_{\rm lib}=4\pi r_{\rm p}/(3\Omega x_{\rm s})$ is the liberation timescale of gas in the disk, $q=M_{\rm p}/M_{*}$, and $q_{\rm d}=\pi r_{\rm p}^2\Sigma/M_{*}$. For the case of slow migration (i.e., $\tau_{\rm lib}\ll 4\pi r_{\rm p}/(3\Omega x_{\rm s})\ll r_{\rm p}/v_{\rm p}$), we were able to assume a steady state for equation \ref{eq:Velocity} to determine $v_{\rm p}$, or in other words:
\begin{equation}
 -v_{\rm p}+\frac{2q_{\rm d}}{\pi qr_{\rm p}^3\Omega \Sigma}(\Gamma_{\rm static}+\Gamma_{\rm dynamic})=0.
\end{equation}

\subsection{Inviscid case: $\frac{x_{\rm s}^2}{6r_{\rm p}\nu}v_{\rm p}>1$}
We derived $v_{\rm p}$ by substituting equations \ref{eq:TorqueDynamic} and \ref{eq:Vortencity} into equation \ref{eq:Velocity} as:
\begin{equation}
\Gamma_{\rm inviscid}=\frac{1}{1-(3/2+p)m_{\rm c}}\Gamma_{\rm static},
\end{equation}
where, $m_{\rm c}$ is given by (Pa14 eq20): 
\begin{equation}
m_{\rm c}=4q_{\rm d}\bar{x}_{\rm s}/q,
\end{equation}
where $\bar{x}_{\rm s}=x_{\rm s}/r_{\rm p}$.

\subsection{Viscid case: $\frac{x_{\rm s}^2}{6r_{\rm p}\nu}v_{\rm p}<1$}
We derived a quadratic equation of $v_{\rm p}$ by substituting equations \ref{eq:TorqueDynamic} and \ref{eq:Vortencity} into equation \ref{eq:Velocity} as:
\begin{equation}
Av_{\rm p}^2 +Bv_{\rm p} +C=0,
\end{equation}
where
\begin{eqnarray}
A&=&\frac{2q_{\rm d}\bar{x}_{\rm s}^3r_{\rm p}}{3q\nu}\left(\frac{3}{2}+p\right)=m_{\rm c}\left(\frac{3}{2}+p\right)\frac{\tau_{\nu}}{6r_{\rm p}}\\
B&=&-1\\
C&=&\frac{2q_{\rm d}q}{\pi h^2}r_{\rm p}\Omega\gamma_{static}=\frac{r_{\rm p}}{\tau_{\rm mig}}\gamma_{\rm static},
\end{eqnarray}
where $h$ is the scale height of the disk, $\gamma_{\rm static}=\Gamma_{\rm static}/\Gamma_0$ ($\Gamma_0 =(q/h)^2\Sigma r^4\Omega^2$), and $\tau_{\nu}$ and $\tau_{\rm mig}$ are the timescales of diffusion and migration, respectively, as given by (Pa14 eq10 and 23):
\begin{eqnarray}
\tau_{\nu}&=&\frac{r_{\rm p}^2\bar{x}_{\rm s}^2}{\nu} \\
\tau_{\rm mig}&=&\frac{\pi h^2}{2q_{\rm d}q\Omega}
\end{eqnarray}

The quadratic formula gives the total torque after dynamic correction:
\begin{equation}
\Gamma_{\rm viscid}=\Theta (k)\Gamma_{\rm static},
\end{equation}
where the function $\Theta(k)$ is defined by (Pa14 eq30):
\begin{equation}
\Theta(k)=\frac{1-\sqrt{1-2k}}{k},
\end{equation}
where $k$ is the coefficients given by (Pa14 eq31-32):
\begin{equation}
k=\frac{8}{3\pi}\left(\frac{3}{2}+p\right)\frac{\gamma_{\rm static} q_{\rm d}^2\bar{x}_{\rm s}^3}{h^2}\frac{r_{\rm p}\Omega}{\nu}
=\left(\frac{3}{2}+p\right)\frac{m_{\rm c}\tau_{\nu}\gamma_{\rm static}}{6\tau_{\rm mig}}.
\end{equation}

The function $\Theta(k)$ takes a critical value of 2 at $k=1/2$, but for $k>1/2$ it does not take any value, since the inside of the square root of $\Theta(k)$ becomes negative. This suggests that runaway migration takes place for the case $k>1/2$. \cite{paa14} suggested that the time scale of migration for runaway case would be $m_{c}\tau_{mig}$, and in such a case, torque for the runaway migration would be:
\begin{equation}
\Gamma_{\rm rw}=\frac{q}{4q_{\rm d}}\Gamma_{0}
\end{equation}

The results are consistent with the numerical results of \cite{paa11}. In their simulation, $v_{\rm p}$ rapidly converged to the values obtained here, after a short transient phase (Figs. 7, 8, and 9 in \cite{paa14}).

In summary, our new torque formula of Type I migration, taking into account dynamic effects, is given as:
\begin{eqnarray}
\Gamma_{I}&=&\Gamma_{\rm static}\min\left(\frac{1}{1-m_{\rm c}(2/3-\alpha)}, \Theta(k)\right) \quad k<0.5
\label{eq:Theta}
\\
&=&\frac{q}{4q_{\rm d}}\Gamma_{0}\quad k>0.5
\end{eqnarray}

\begin{figure}[p]
\begin{minipage}{0.5\hsize}
\includegraphics[width=80mm]{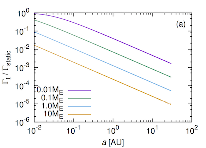}
\end{minipage}
\begin{minipage}{0.5\hsize}
\includegraphics[width=80mm]{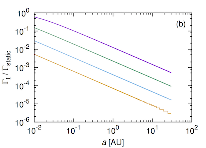}
\end{minipage}
\begin{minipage}{0.5\hsize}
 \includegraphics[width=80mm]{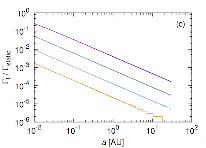}
\end{minipage}
\begin{minipage}{0.5\hsize}
 \includegraphics[width=80mm]{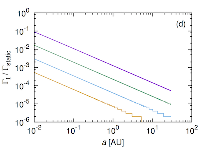}
\end{minipage}
\begin{center}
    \caption{Dynamic correction factor at each semimajor axis for 0.01 $M_{\oplus}$ ($M_{\oplus} = M_E$), 0.1 $M_{\oplus}$, 1.0 $M_{\oplus}$, and 10 $M_{\oplus}$ from top to bottom in each panel. The mass of the disks are (a) 1/$\sqrt{10} \times$MMSN, (b) 1.0$\times$MMSN, (c) $\sqrt{10}\times$MMSN, and (d) 10$\times$MMSN.}
  \end{center}
\end{figure}

Figure 1 shows the dynamic correction factor at each semimajor axis for embryos with masses of 0.01 $M_{\oplus}$, 0.1 $M_{\oplus}$, 1.0 $M_{\oplus}$, and 10 $M_{\oplus}$. The masses of the disks were (a) 1/$\sqrt{10} \times$MMSN, (b) 1.0$\times$MMSN, (c) $\sqrt{10}\times$MMSN, and (d) 10$\times$MMSN. Except for a close-in small protoplanet, most protoplanets had correction factors of significantly less than 0.1; therefore, Type I migration was generally significantly slowed by dynamic effects.

\section{Planet Formation and Migration Model}

In our model, we adopted the models of \cite{ida04, ida08} and \cite{ida13} for (1) planetesimals' growth through cohesive collisions, (2) the evolution of planetesimal surface density, (3) embryos' Type I migration and their stoppage at the disk inner edge (except for a modification of the Type I migration formula to include dynamic correction; see Section 2), and for the gas giants, (4) the onset, rate, and termination (through gap opening and/or global depletion) of efficient gas accretion, and (5) their Type II migration.

\subsection{Disk models}

We adopted the MMSN model \citep{hay81} as a fiducial set of initial conditions for planetesimal surface density ($\Sigma _d$) and introduced a multiplicative factor ($f_d$). For the gas surface density ($\Sigma _g$), we adopted the $r$-dependence of steady accretion disk with constant viscosity ($\Sigma _g \propto r^{-1}$) scaled by that of the MMSN at 10 AU with a scaling factor ($f_g$). Following \cite{ida08}, we set:
\begin{equation}
\left\{ \begin{tabular}{@{}l@{}}
$\Sigma_d = \Sigma_{d,10}\eta_{\rm{ice}}f_d(r/10\rm{AU})^{-1.5}$,\\
$\Sigma_g = \Sigma_{g,10}f_g(r/10\rm{AU})^{-1.0}$,
\end{tabular}\right.
\end{equation}
where normalization factors $\Sigma_{d,10} = 0.32$ g cm$^{-2}$ and $\Sigma_{g,10} = 75$ g cm$^{-2}$, and the step function was $\eta_{\rm{ice}} = 1$ inside the ice line at $a_{\rm{ice}}$ and 4.2 for $r > a_{\rm{ice}}$.

Neglecting the detailed energy balance in the disk \citep{chi97}, we adopted the equilibrium temperature distribution of optically thin disks given by \cite{hay81}, such that:
\begin{equation}
T = 280\left(\frac{r}{1\rm{AU}} \right)^{-1/2}\left(\frac{L_*}{L_{\odot}} \right)^{1/4} \rm{K},
\end{equation}
where $L_*$ and $L_{\odot}$ are stellar and solar luminosity. We set the ice line to be that determined by an equilibrium temperature in optically thin disk regions:
\begin{equation}
a_{\rm{ice}} = 2.7\left(\frac{L_*}{L_{\odot}} \right)^{1/2} \rm{AU}.
\end{equation}

Owing to viscous diffusion and photoevaporation, $f_g$ decreases with time. For simplicity, we adopted:
\begin{equation}
f_g = f_{g,0}\exp\left(-\frac{t}{\tau_{\rm{dep}}} \right),
\end{equation}
where $f_{g,0}$ is the initial value of $f_g$ and $\tau_{\rm{dep}}$ is the gas depletion timescale.

\subsection{From oligarchic growth to isolation}

On the basis of the oligarchic growth model \citep{kok98, kok02}, the growth rate of embryos/cores at any location, $a$, and time $t$, in the presence of disk gas, was described by:
\begin{equation}
\frac{dM_c}{dt} = \frac{M_c}{\tau_{\rm{c,acc}}}
\end{equation}
where
\begin{equation}
\tau_{\rm{c,acc}} = 3.5\times 10^5 \eta_{\rm{ice}}^{-1}f_d^{-1}f_g^{-2/5}\left(\frac{a}{1\rm{AU}} \right)^{5/2}\left(\frac{M_c}{M_{\oplus}} \right)^{1/3}\left(\frac{M_*}{M_{\odot}} \right)^{-1/6} \rm{yr},
\end{equation}
where $M_c$ is the mass of the embryo/core. Furthermore, we set the mass of typical field planetesimals to be $m = 10^{20}$ g.

We computed the evolution of $\Sigma_d$ distribution due to accretion by all emerging embryos in a self-consistent manner. The growth and migration of many planets were integrated simultaneously with the evolution of the $\Sigma_d$-distribution.

During the early phase of evolution, embryos are embedded in their natal disks. Despite their mutual gravitational perturbation, embryos preserve their circular orbits owing to gravitational drag from disk gas \citep{war93} and dynamic friction from residual planetesimals \citep{ste00}. After the disk gas is severely depleted, the efficiency of the eccentricity damping mechanism is reduced, and the embryos' eccentricity grow until they cross each other's orbits (i.e., giant impact). However, in this study, growth via the giant impact process was not considered. Moreover, we also ignored dynamic interaction between planets, with the growth of individual planets integrated independently.

\subsection{Type I migration}

Type I migration of an embryo is caused by the sum of tidal torque from disk regions that are both interior and exterior to the embryos. The rate and direction of embryos' migration are determined by the differential Lindblad and corotation torques. While a conventional formula of Type I migration, which assumes locally isothermal disks \citep{tan02}, shows that the migration is always inward, recent developments have shown Type I migration of isolated embryos in non-isothermal disks; therefore, the magnitude and sign of tidal torque can be changed. \cite{ida08} used the conventional formula of Type I migration in isothermal disks derived by \cite{tan02} with a scaling factor $C_1$ of:
\begin{equation}
\frac{dr}{dt} \simeq C_1 \times 1.08(p + 0.80q -2.52)\frac{M_p}{M_*}\frac{\Sigma_g r^2}{M_*}\left(\frac{r\Omega_{\rm{K}}}{c_s} \right)^2 r\Omega_{\rm{K}},
\end{equation}
where $p = d\log \Sigma_g / d \log r$, $q = d \log T / d \log r$, $c_s$ is the sound speed, and $\Omega_{\rm{K}}$ is the Keplerian angular velocity. The expression of \cite{tan02} corresponds to $C_1 = 1$, and for slower migration, $C_1 < 1$. While we derived a new torque formula for Type I migration that included dynamic corrections (see Section 2), for comparison we also used the conventional formula with the scaling factor $C_1 = 1.0$.

We assumed that Type I migration ceases inside the inner boundary of the disk, because at this point $f_g$ is locally zero. For computational convenience, we set the disk inner boundary to be the edge of the magnetospheric cavity at 0.04 AU.

\subsection{Formation of gas giant planets}

Models for the formation of gas giant planets were the same as those used in \cite{ida13}. Embryos were surrounded by gaseous envelopes when their surface escape velocities became larger than the sound speed of the surrounding disk gas. When their mass grew (through planetesimal bombardment) above a critical mass:
\begin{equation}
M_{c,\rm{hydro}} \simeq 10\left(\frac{\dot{M}_c}{10^{-6}M_{\oplus}\rm{yr}^{-1}} \right)^{0.25}M_{\oplus},
\end{equation}
both the radiative and convective transport of heat became sufficiently efficient to allow their envelope to contract dynamically \citep{iko00}.

In the above equation, we neglected the dependence on opacity in the envelope \citep{hor10}. In regions where cores have already acquired isolation mass, their planetesimal-accretion rate ($\dot{M}_c$) would be much diminished \citep{iko00} and $M_{c,\rm{hydro}}$ would be comparable to an Earth-mass, $M_{\oplus}$. However, gas accretion also releases energy and its rate is still regulated by the efficiency of radiative transfer in the envelope, such that:
\begin{equation}
\frac{dM_p}{dt} \simeq \frac{M_p}{\tau_{\rm{KH}}},
\label{eq:Gasaccretion}
\end{equation}
where $M_p$ is the planet mass including gas envelope. According to \cite{ida08}, we approximated the Kelvin-Helmholtz contraction timescale, $\tau_{\rm{KH}}$, of the envelope using:
\begin{equation}
\tau_{\rm{KH}} \simeq 10^9\left(\frac{M_p}{M_{\oplus}} \right)^{-3},
\end{equation}

Equation \ref{eq:Gasaccretion} shows that $dM_p/dt$ rapidly increases as $M_p$ grows; however, this is limited by the global gas accretion rate throughout the disk and by the process of gap formation near the protoplanets' orbits. The disk accretion rate can be expressed as:
\begin{equation}
\dot{M}_{\rm{disk}} \simeq 3 \times 10^{-9} f_g \left(\frac{\alpha}{10^{-3}} \right) M_{\rm{\odot}} \rm{yr}^{-1},
\end{equation}
where $\alpha$ is a parameter of alpha prescription for turbulent viscosity \citep{sha73}. During the advanced stage of disk evolution, we assumed that both $\dot{M}_{\rm{disk}}$ and $\Sigma_g$ evolved in proportion to $\exp(-t/\tau_{\rm{dep}})$. The rate of accretion onto the cores cannot exceed $\dot{M}_{\rm{disk}}$.

A gap, or at least a partial gap, is formed when a planet's tidal torque exceeds the disk's intrinsic viscous stress \citep{lin85}. This viscous condition for gap formation is satisfied for planets with:
\begin{equation}
M_p > M_{g,\rm{vis}} \simeq 30\left(\frac{\alpha}{10^{-3}} \right)\left(\frac{a}{1\rm{AU}} \right)^{1/2}\left(\frac{L_*}{L_{\odot}} \right)^{1/4} M_{\oplus}.
\end{equation}
In this case, Type I migration transitions to Type II migration. Fluid dynamic simulations \citep{dan03, dan08} show that some fraction of gas still flows into the gap. Following the results of \cite{dob07}, we completely terminated gas accretion when a planet's Hill radius became larger than two times the disk scale height, which corresponded to the thermal condition of \citep{lin85}, that is:
\begin{equation}
M_p > M_{g,\rm{th}} \simeq 0.95 \times 10^3 \left(\frac{a}{1\rm{AU}} \right)^{3/4}\left(\frac{L_*}{L_{\odot}} \right)^{3/8}\left(\frac{M_*}{M_{\odot}} \right)^{-1/2} M_{\oplus}.
\end{equation}

In general, our models for gas accretion rates onto the cores were:
\begin{equation}
\frac{dM_p}{dt} = f_{\rm{gap}} \dot{M}_{p\rm{,nogap}},
\end{equation}
when in the absence of any feedback on the disk structure. Therefore, without the effect of gap opening:
\begin{equation}
\dot{M}_{p\rm{,nogap}} =  \min \left(\frac{M_p}{\tau_{\rm{KH}}}, \dot{M}_{\rm{disk}} \right),
\end{equation}
and $f_{\rm{gap}}$ is a reduction factor due to gap opening:
\begin{equation}
f_{\rm{gap}} = 
\left\{ \begin{tabular}{@{}l@{}}
$1$ \;\;\;\;\;\;\;\;\;\;\;\;\;\;\;\;\;\;\;\;\;\;\;\;\;\;\;\;\;\;\;\;\;\;\;\:[for $M_p < M_{\rm{g,vis}}$]\\
$\frac{\displaystyle \log M_p - \log M_{\rm{g,vis}}}{\displaystyle \log M_{\rm{g,th}} - \log M_{\rm{g,vis}}}$ \;\;\;\;\;[for $M_{\rm{g,vis}} < M_p < M_{\rm{g,th}}$],\\
$0$ \;\;\;\;\;\;\;\;\;\;\;\;\;\;\;\;\;\;\;\;\;\;\;\;\;\;\;\;\;\;\;\;\;\;\;\:[for $M_p > M_{\rm{g,th}}$].
\end{tabular}\right.
\end{equation}

\subsection{Type II migration}

During gap formation, embedded gas giants adjust their positions in the gap to establish a quasi equilibrium between the torque applied on them from disk regions both interior and exterior to their orbits. Subsequently, as the disk gas undergoes viscous diffusion, this interaction leads to Type II migration.

We assumed that planets undergo Type II migration after they have accreted a sufficient mass to satisfy the viscous condition ($M_{g,\rm{vis}} < M_p$) for gap formation.

While $M_p$ increases, the disk mass declines owing to stellar and planetary accretion and photoevaporation. While disk mass exceeds $M_p$ (during the disk-dominated regime), planets' Type II migration is locked, with the viscous diffusion of the disk gas. During the advanced stages of disk evolution, when the mass becomes smaller than $M_p$ (during the planet-dominated regime), embedded planets carry a major share of the total angular momentum content.

For the disk-dominated regime, the migration timescale is given by:
\begin{equation}
\tau_{\rm{mig2,disk}} \simeq 0.7 \times 10^5 \left(\frac{\alpha}{10^{-3}} \right)^{-1}\left(\frac{a}{1\rm{AU}} \right)\left(\frac{M_*}{M_{\odot}} \right)^{-1/2} \rm{yr}.
\end{equation}

For the planet-dominated regime, the migration timescale is given by:
\begin{equation}
\tau_{\rm{mig2,pl}} \simeq 5 \times 10^5 f_g^{-1} \left(\frac{C_2\alpha}{10^{-4}} \right)^{-1}\left(\frac{M_p}{M_J} \right)\left(\frac{a}{1\rm{AU}} \right)^{1/2}\left(\frac{M_*}{M_{\odot}} \right)^{-1/2} \rm{yr},
\end{equation}
where $C_2$ is an efficiency factor associated with the degree of asymmetry in the torques between the inner and outer disk regions. If the inner disk is severely depleted, $C_2 = 1$. We treated the factor $C_2$ as a model parameter, and we set $C_2 = 0.1$.

\section{Population Synthesis of Planetary Systems}

Using our new torque formula for Type I migration, we modeled the formation of planetary systems using Monte Carlo simulations. The predicted mass and period distributions were compared with those from a conventional Type I migration model.

\subsection{Numerical settings}

We first generated a set of 1000 disks with various values of $f_{g,0}$ (the initial value of $f_g$) and $\tau_{\rm{dep}}$. We adopted a range of disk model parameters that represented the observed distribution of disk properties and assigned them to each model with an appropriate statistical weight. For the gaseous component, we assumed that $f_{g,0}$ had a lognormal distribution centered on $f_{g,0} = 1$ with a dispersion of 1 and an upper cutoff at $f_{g,0} = 30$, independent of the stellar metallicity. For heavy elements, we choose $f_{d,0} = 10^{\rm{[Fe/H]}_d}f_{g,0}$, where [Fe/H]$_d$ is the metallicity of the disk. We assumed that these disks had the same metallicities as their host stars. We also assumed that $\tau_{\rm{dep}}$ had log-uniform distributions in the range $10^6--10^7$ yr.

For each disk, 15 values of a for the protoplanetary seeds were selected from a long-uniform distribution in the range 0.05--30 AU, assuming that the mean orbital separation between planets was 0.2 on a logarithmic scale. Constant spacing in the logarithm corresponded to the spacings between the cores, which were proportional to $a$. This represented the simplest choice and a natural outcome of dynamic isolation at the end of oligarchic growth.

In all simulations, the values $\alpha = 10^{-3}$ and $M_* = 1M_{\odot}$ were assumed. Since ongoing radial velocity surveys are focused on relatively metal-rich stars, we presented our results with [Fe/H] = 0.1.

We artificially terminated Type I and Type II migration near the disk inner edge at 0.04 AU. We did not specify a survival criterion for the close-in planets because we lacked adequate knowledge about planets' migration and their interaction with host stars near the inner edge of their nascent disks. Hence, we recorded all of the planets that migrated to the vicinity of their host stars. In reality, a large fraction of the giant planets that have migrated to small disk radii were either consumed \citep{san98} or tidally disrupted \citep{tri98} by their host stars. Cores that migrate to the inner edge of the disk may also coagulated and form super-Earths \citep{ogi09}; however, this was not considered in our simulations.

\subsection{Simulated individual systems}

\begin{figure}[p]
\begin{minipage}{0.5\hsize}
  \includegraphics[width=80mm]{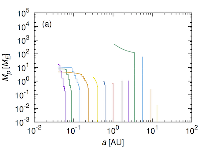}
 \end{minipage}
 \begin{minipage}{0.5\hsize}
  \includegraphics[width=80mm]{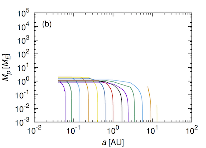}
\end{minipage}
 \begin{center}
  \caption{Growth and migration of planets for scaling factor $f_{g,0} = 3.0$. Units of mass ($M_p$) and semimajor axis ($a$) are Earth masses ($M_{\oplus} = M_E$) and AU. (a) Mass evolutions obtained from simulations with the new torque formula for Type I migration. (b) Evolutions using the conventional formula ($C_1 = 1.0$).}
  \end{center}
\end{figure}

\begin{figure}[p]
\begin{minipage}{0.5\hsize}
  \includegraphics[width=80mm]{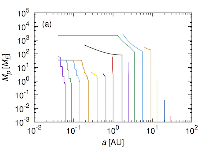}
 \end{minipage}
 \begin{minipage}{0.5\hsize}
  \includegraphics[width=80mm]{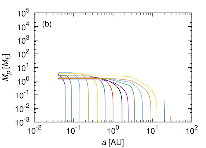}
 \end{minipage} 
  \begin{center}
    \caption{Growth and migration of planets for scaling factor $f_{g,0} = 8.0$. Units of mass ($M_p$) and semimajor axis ($a$) are Earth masses ($M_{\oplus} = M_E$) and AU. (a) Mass evolutions obtained from simulations with the new torque formula for Type I migration. (b) Evolutions using the conventional formula ($C_1 = 1.0$).}
  \end{center}
\end{figure}

\begin{figure}[h]
\begin{minipage}{0.5\hsize}
  \includegraphics[width=80mm]{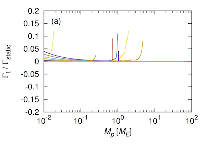}
 \end{minipage}
 \begin{minipage}{0.5\hsize}
  \includegraphics[width=80mm]{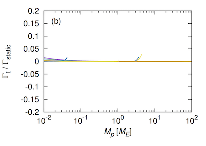}
 \end{minipage}
  \begin{center}
   \caption{Evolution of the dynamic correction factor for a scaling factor of (a) $f_{g,0} = 3.0$, and (b) $f_{g,0} = 8.0$. Units of mass ($M_p$) are Earth masses ($M_{\oplus} = M_E$).}
 \end{center}
\end{figure}

\begin{figure}[h]
 \begin{minipage}{0.5\hsize}
  \includegraphics[width=80mm]{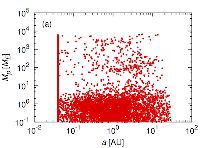}
 \end{minipage}
 \begin{minipage}{0.5\hsize}
  \includegraphics[width=80mm]{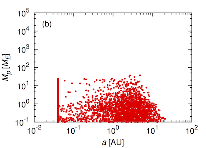}
 \end{minipage}
 \begin{minipage}{\hsize}
 \begin{center}
  \includegraphics[width=80mm]{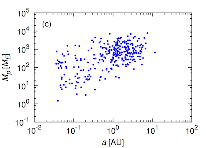}
 \end{center}
 \end{minipage}
  \begin{center}
   \caption{Planetary mass and semimajor axis distribution. Units of mass ($M_p$) and semimajor axis ($a$) are Earth masses ($M_{\oplus} = M_E$) and AU. (a) Distribution obtained from Monte Carlo simulations with the new torque formula for Type I migration. (b) Distributions using the conventional formula ($C_1 = 1.0$). (c) Observational data of extrasolar planets around stars with $M_* = 0.8-1.2 M_{\odot}$ detected by radial velocity surveys. The determined value of $M_p\sin i$ is multiplied by $1/\average{\sin i} = 4/\pi \simeq 1.27$, where a random orientation of the planetary orbital planes is assumed.}
  \end{center}
\end{figure}

We compared the time evolution of planetary masses and semimajor axes for the new torque formula model (Fig. 2a; Fig. 3a) and the conventional torque formula model with $C_1 = 1.0$ (Fig. 2b; Fig. 3b). We choose a disk a few times more massive ($f_{g,0} = 6.0$ and $8.0$) than the minimum solar nebula, and with $\tau_{\rm{dep}} = 3\times 10^6$ yr. The results showed that inward migration of planet embryos was slowed significantly by dynamic effects. When dynamic effects were considered, some cores survived and grew into gas giants; however, when considering only static torque, all cores migrated to the vicinity of their central star before growing enough to accrete the nebula gas.

Figure 4 shows the evolution of the dynamic correction factor (see equation \ref{eq:Theta}),
\begin{equation}
\Gamma_I / \Gamma_{\rm{static}} = \min\left(\frac{1}{1-m_{\rm c}(3/2 + p)}, \Theta(k)\right),
\end{equation}
with the mass of each planet embryo. The correction factors remained small ($\leq$ 0.1) throughout the simulation; therefore, the dynamic correction of Type I migration effectively prevented embryos from migrating to the central star.

\subsection{Distributions of mass and semimajor axes}

We compared the predicted $M_p$-$a$ distributions using the new torque formula (Fig. 5a) and the conventional formula with $C_1 = 1.0 $ at $t = 2\times 10^7$ yr. In order to directly compare the theoretical predictions with the observed data, we plotted values of $M_p$ that were 1.27 times the values of $M_p\sin i$, as determined from radial velocity measurements (Fig. 5c). This correction factor corresponded to mean values of $1/\average{\sin i} = 4/\pi$ for a sample of planetary systems with randomly oriented orbital plants. To compare the theoretical results with $M_* = 1M_{\odot}$, we plotted only the data of planets around stars with $M_* = 0.8-1.2 M_{\odot}$ that have been observed by radial velocity surveys \footnote{See http://exoplanet.eu/.}.

For the conventional models, the formation probability of gas giants dramatically changed with $C_1$ \citep{ida08}. Within the limits of Type I migration with an efficiency comparable to that deduced from the traditional linear torque analysis (i.e., with $C_1 = 1$; Fig. 5b), all cores were cleared prior to gas depletion, such that gas giant formation was effectively suppressed. However, when considering the dynamic effects, a substantial fraction of the cores survived and grew into large gas giants (Fig. 5a). We carry out a Kolmogorov-Smirnov (K-S) test for statistical similarity between the predicted $M_p$-$a$ distributions and the observed data for the parameter domain of 0.1 AU $<$ a $<$ 5 AU and $M_p > 100 M$. While the conventional model produces no giant planets (Fig. 5b), the predicted $M_p$-$a$ distribution using the new torque formula (Fig. 5a) is statistically similar to the observed data (Fig. 5c) within a significance level of $p$-value $> 0.05$ for both the semimajor axis and mass cumulative distribution functions.

Without considering the dynamic correction for Type I migration, when a planet's mass exceeded that of Earth, the corotation torque became smaller owing to saturation. At this point, static torque affected the planet more efficiently so that its inward migration was rapid. However, when the dynamic correction was included, the migration timescale was short, and the saturation of the corotation torque was less effective. Under these conditions, inward migration slowed, which allowed for the formation of gas giants before migration to the central star.

In summary, population synthesis simulations using our new torque formula with dynamic correction (Fig. 5a) can explain the gas giants ($> 100 M_{\oplus}$) observed in exoplanetary systems (Fig. 5c). In contrast, simulations using a conventional formula ($C_1 = 1.0$; Fig. 5b) cannot explain the observed data. These results show that planet populations consistent with observations can be reproduced naturally (i.e., without considering the reduction factor) if we take into account dynamic corrections for Type I migration torque.

\section{Conclusions}

We derived a new torque formula for Type I migration by taking into account dynamic corrections. Using this formula, we performed population synthesis simulations with and without the effects of dynamic torques. In most cases, inward migration was significantly slowed by the dynamic effects. Considering just static torques, gas giant formation was effectively suppressed by Type I migration of cores; however, when dynamic effects were considered, a substantial fraction of cores survived and grew into gas giants.

\section*{Acknowledgments}
We thank an anonymous reviewer for the helpful comments. This research was supported by Grant-in-Aid for Scientific Research on Innovative Areas from the Ministry of Education, Culture, Sports, Science and Technology (MEXT; Grant Number 26106006). T.S. was supported by a Grand-in-Aid for Young Scientists (KAKENHI B) from the Japan Society for the Promotion of Science (JSPS; Grant Number 24740120).

\appendix

\section{Static Torque Formula}

Static torque (Pa11 and CN14) is given by
\begin{eqnarray}
\Gamma_{\rm static}&=& F_{\rm L}\Gamma_{\rm LR}\\
&+&\left[
\Gamma_{\rm VHS} F(p_\nu)G(p_\nu)
+\Gamma_{\rm EHS}F(p_\nu)F(p_\chi)\sqrt{G(p_\nu)G(p_\chi)}\right.\\
&+&\Gamma_{\rm LVCT}(1-K(p_\nu))
\left.+\Gamma_{\rm LECT}\sqrt{(1-K(p_\nu))(1-K(p_\chi))}
\right]
F_{\rm e}F_{\rm i},
\end{eqnarray}
where $\Gamma_{\rm LR}$, $\Gamma_{\rm VHS}$, $\Gamma_{\rm EHS}$, $\Gamma_{\rm LVCT}$, and $\Gamma_{\rm LECT}$ are the Lindblad torque, vortensity and entropy related horseshoe drag torques, and linear vortensity and entropy related corotation torques, respectively, as given by equation 3--7 in \cite{paa11}:
\begin{eqnarray}
\Gamma_{\rm LR}&=&(-2.5-1.7\beta+0.1\alpha)\Gamma_0/\gamma_{\rm eff},\\
\Gamma_{\rm VHS}&=&[1.1(3/2-\alpha)]\Gamma_0/\gamma_{\rm eff},\\
\Gamma_{\rm EHS}&=&7.9(\xi/\gamma_{\rm eff})\Gamma_{0}/\gamma_{\rm eff},\\
\Gamma_{\rm LVCT}&=&[0.7(3/2-\alpha)]\Gamma_0/\gamma_{\rm eff},\\
\Gamma_{\rm LECT}&=&[(2.2-1.4/\gamma_{\rm eff})]\Gamma_0/\gamma_{\rm eff},
\end{eqnarray}
where $\alpha=d\ln \Sigma/d\ln r$, $\beta=d\ln T_{\rm m}/d\ln r$, and $\xi=\beta-(\gamma_{\rm eff}-1)\alpha$. Here, $\Gamma_0=(q/h)^2\Sigma r^4\Omega^2$. 

The functions $F(p_\nu)$, $F(p_\chi)$, $G(p_\nu)$, $F(p_\chi)$, $K(p_\nu)$, and $K(p_\chi)$ are related to the ratio between the viscous/thermal diffusion time scale and horseshoe liberation/horseshoe U-turn time scales, given by equations 23, 30, and 21 in \cite{paa11}:

\begin{equation}
F(p)=\frac{1}{1+(p/1.3)^2}
\end{equation}

\begin{eqnarray}
G(p)=&\frac{16}{25}\left(\frac{45\pi}{8}\right)^{3/4}p^{3/2}\quad p<\sqrt{\frac{8}{45\pi}}\\
&1-\frac{9}{25}\left(\frac{8}{45\pi}\right)^{4/3}p^{-8/3}\quad p>\sqrt{\frac{8}{45\pi}}
\end{eqnarray}

\begin{eqnarray}
K(p)=&\frac{16}{25}\left(\frac{45\pi}{28}\right)^{3/4}p^{3/2}\quad p<\sqrt{\frac{28}{45\pi}}\\
&1-\frac{9}{25}\left(\frac{28}{45\pi}\right)^{4/3}p^{-8/3}\quad p>\sqrt{\frac{28}{45\pi}}.
\end{eqnarray}

The $p_\nu$ and $p_\chi$ are given by (Pa11 eq19 and eq40):
\begin{eqnarray}
p_{\nu}&=&\frac{2}{3}\sqrt{\frac{r^2\Omega \bar{x}_{\rm s}^3}{2\pi\nu}}\\
p_{\chi}&=&\sqrt{\frac{r^2\Omega \bar{x}_{\rm s}^3}{2\pi\chi}},
\end{eqnarray}
where (Pa11 eq48-49) $\bar{x}_{\rm s}=x_{\rm s}/r$ is given by
\begin{equation}
\bar{x}_{\rm s}=C\sqrt{q/h},
\end{equation}
\begin{equation}
C=\frac{1.1}{\gamma_{\rm eff}^{1/4}}\left(\frac{0.4}{b/h}\right)^{-1/4},
\end{equation}
and (Pa11 eq34)
\begin{equation}
\chi=\frac{4\gamma(\gamma-1)\sigma T^4}{3\kappa\rho^2H^2\Omega^2}.
\end{equation}

The effective adiabatic index $\gamma_{\rm eff}$ is given by (Pa11 eq45-46):
\begin{equation}
\gamma_{\rm eff}=\frac{
2Q\gamma}
{
\gamma Q+\frac{1}{2}\sqrt{2\sqrt{(\gamma^2 Q^2+1)2-16Q^2(\gamma-1)}+2\gamma^2Q2-2}
},
\end{equation}
taking into account of the photon diffusion in a disk. Here,
\begin{equation}
Q=\frac{2\chi}{3h^3r^2\Omega}
\end{equation}

$F_{\rm e}$ and $F_{\rm i}$ are the reduction factor due to eccentricity and inclination of the planets, which are give by (CN15 eq16-20):
\begin{equation}
F_{\rm e}=\exp\left(-\frac{e}{e_{\rm f}}\right),
\end{equation}
where $e$ is the plane's eccentricity and $e_{\rm f}$ is defined as: 

\begin{equation}
e_{\rm f}=h/2+0.01
\end{equation} 

\begin{equation}
F_{\rm i}=1-\tanh(i/h),
\end{equation}
where $i$ is the inclination of the planet. The factor $F_{\rm L}$ is the reduction in Lindblad torques when planets are on eccentric or inclined orbits, and is given by \cite{cre08}:

\begin{equation}
F_{\rm L}=\left[P_{\rm e}+\frac{P_{\rm e}}{|P_{\rm e}|}\times 
\left( 
0.07\left(\frac{i}{h}\right)+
0.085\left(\frac{i}{e}\right)^4-0.08\left(\frac{e}{h}\right)\left(\frac{i}{h}\right)^2
\right)\right]^{-1},
\end{equation}
where $P_{\rm e}$ is defined as 
\begin{equation}
P_{\rm e}=\frac{1+\left(\frac{e}{2.25h}\right)^{1/2}+\left(\frac{e}{2.84h}\right)^6}{1-\left(\frac{e}{2.02h}\right)^4}
\end{equation}

\end{document}